\documentclass[reprint,amsmath,amssymb,aps,twocolumn,superscriptaddress]{revtex4-2}
\usepackage{graphicx}
\usepackage{dcolumn}
\usepackage{bm}
\usepackage{float}
\usepackage{amssymb}
\usepackage{color}
\usepackage{multirow}
\usepackage{stmaryrd}
\usepackage[colorlinks,linkcolor=blue,urlcolor=blue,citecolor=blue]{hyperref}

\begin{document}


\title{Pump-induced terahertz conductivity response and peculiar bound state in Mn$_3$Si$_2$Te$_6$}

\author{Qiong Wu}
\email{qwu20@pku.edu.cn}
\affiliation{International Center for Quantum Materials, School of Physics, Peking University, Beijing 100871, China}

\author{Qiangwei Yin}
\affiliation{Department of Physics and Beijing Key Laboratory of Opto-electronic Functional Materials $\&$ Micro-nano Devices, Renmin University of China, Beijing 100872, China}

\author{Sijie Zhang}
\affiliation{International Center for Quantum Materials, School of Physics, Peking University, Beijing 100871, China}

\author{Tianchen Hu}
\affiliation{International Center for Quantum Materials, School of Physics, Peking University, Beijing 100871, China}

\author{Dong Wu}
\affiliation{Beijing Academy of Quantum Information Sciences, Beijing 100193, China}

\author{Li Yue}
\affiliation{International Center for Quantum Materials, School of Physics, Peking University, Beijing 100871, China}

\author{Bohan Li}
\affiliation{Beijing Academy of Quantum Information Sciences, Beijing 100193, China}

\author{Shuxiang Xu}
\affiliation{International Center for Quantum Materials, School of Physics, Peking University, Beijing 100871, China}

\author{Rongsheng Li}
\affiliation{International Center for Quantum Materials, School of Physics, Peking University, Beijing 100871, China}

\author{Qiaomei Liu}
\affiliation{International Center for Quantum Materials, School of Physics, Peking University, Beijing 100871, China}

\author{Hechang Lei}
\affiliation{Department of Physics and Beijing Key Laboratory of Opto-electronic Functional Materials $\&$ Micro-nano Devices, Renmin University of China, Beijing 100872, China}

\author{Tao Dong}
\affiliation{International Center for Quantum Materials, School of Physics, Peking University, Beijing 100871, China}

\author{Nanlin Wang}
\email{nlwang@pku.edu.cn}
\affiliation{International Center for Quantum Materials, School of Physics, Peking University, Beijing 100871, China}
\affiliation{Beijing Academy of Quantum Information Sciences, Beijing 100193, China}


\begin{abstract}

We report the significant enhancement on ultrafast terahertz optical conductivity and the unexpected formation of a polaronic-like state in semiconductor Mn$_3$Si$_2$Te$_6$ at room temperature. With the absorption of pump photons, the low-frequency terahertz photoconductivity spectrum exhibits a significant rise, quickly forming a broad peak and subsequently shifting to higher energy. The short-lived nature of the broad peak, as well as the distribution of optical constants, strongly points towards a transient polaron mechanism. Our study not only provides profound insights into the remarkable photoelectric response of Mn$_3$Si$_2$Te$_6$ but also highlights its significant potential for future photoelectric applications.

\end{abstract}

\pacs{Valid PACS appear here}
\maketitle


The optical response of condensed matter has been a long-standing research topic. It not only reveals the fundamental physical properties of the materials, but also facilitates the development of high-performance optical devices. In recent years, new possibilities for investigating the non-equilibrium state of condensed matter have emerged with the implementation of the optical pump-probe method. With optical pulses, carriers are excited to the available conduction bands. By monitoring their transient occupation or relaxation dynamics, the band structure information above the Fermi level can be obtained. This technique has been successfully applied to capture the gap evolution of superconductors\cite{Demsar1999,Chia2010,Torchinsky2010,Tian2016}, charge density wave materials\cite{Yusupov2008,Wang2021}, and other systems\cite{Chu2017}. Combined with low frequency terahertz (THz) spectroscopy, the more comprehensive complex dielectric function (i.e. the complex optical conductivity) can be derived to reveal nonequilibrium carrier dynamics and novel excitation modes\cite{Suzuki2012,Kaindl2009}, at specific time points. As a contact-free, energy-resolved and noninvasive technique, optical pump-THz probe spectroscopy plays an increasingly important role in exploring the non-equilibrium carrier dynamics near the Fermi surface and low energy excitation.

Owing to fertile and fanscinating electronic and optical properties, semicondutors, including the transition metal dichalcogenides, have become an integral part of the field of optoelectronic applications\cite{Mak2016}. Mn$_3$Si$_2$Te$_6$, a semiconductor with trigonal structure \cite{Rimet1981,Vincent1986,May2017}, which has recently received considerable attention because of its novel ferrimagnetic state \cite{Seo2021,Ni2021}. It hosts a space group of $P\overline{3}1c$ where the Mn atoms occupy two different positions. Below 78 K, the spin of Mn atoms at two point sites become antiparallel in the ab plane. Due to the different number of Mn atoms between two positions, a ferrimagnetic state is formed. At low temperature, with applying magnetic field along c-axis, the in-plane resistance can be reduced by several orders of magnitude by coupling of chiral orbital currents\cite{Seo2021,Ni2021,Zhang2023}. Except for the colossal magnetoresistance, applying current\cite{Zhang2022} or pressure\cite{Wang2022} can also significantly modulate the magnetism and metallicity of Mn$_3$Si$_2$Te$_6$. All these interesting phenomena indicate that the ground state of Mn$_3$Si$_2$Te$_6$ exhibits complex interactions and is not particularly stable, which can be easily disturbed by external stimulus. Therefore, it is intriguing to explore whether pulsed laser can induce novel states or responses in this system.

In this study, we employ the optical pump-THz probe method to investigate the transient conductivity response induced by optical pumping in the semiconductor Mn$_3$Si$_2$Te$_6$. Upon excitation with a near-infrared (NIR) laser pulse, we observe a significant enhancement in the reflected THz electronic field. The transformed optical conductivity reveals a substantial increase by several orders of magnitude within the skin depth of the pump beam. Unlike the typical Drude response, we observe the main feature as a broad resonance peak at finite energy. The transient conductivity exhibits an increase from 20 to a maximum of $\sim$1000 $\Omega$$^{-1}$ cm$^{-1}$ at a pump fluence of 1.04 mJ/cm$^{2}$. The delay time evolution investigation shows that the resonant mode emergies from low frequency and quickly stabilizes after 4 ps. This resonant state decays subsequently with 2.2 ps lifetime, indicating effective energy transmission induced by strong electron-phonon coupling. By using the polaron response model, the conductivity spectra can be well described and gives binding energy at about 5 meV. Our work unveils a novel photo-induced transient state and highlights the high optical-electronic performance of Mn$_3$Si$_2$Te$_6$.

\begin{figure*}[t]
	\centering
	\includegraphics[width=17cm]{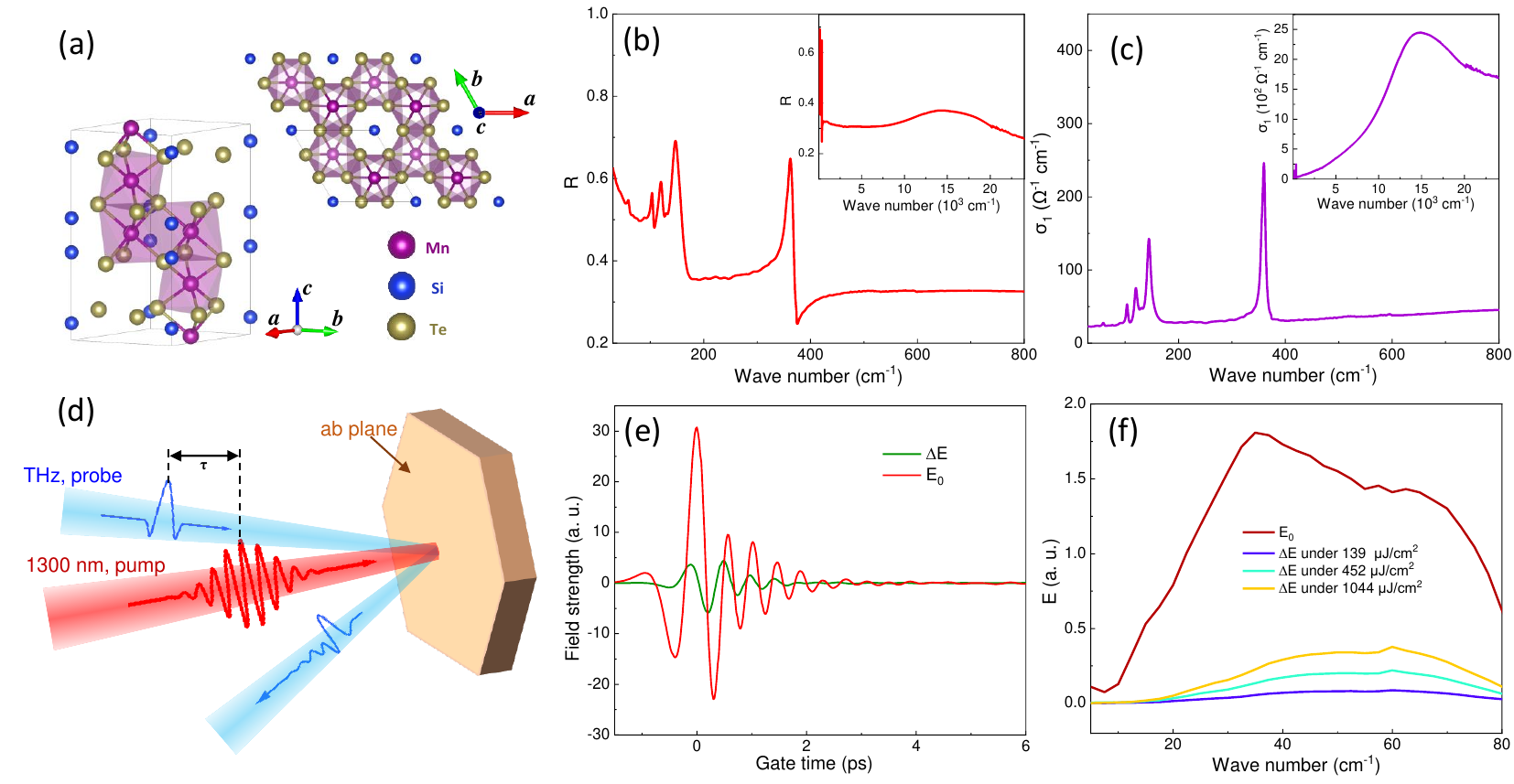}\\
	\caption{\textbf{The optical characterization of Mn$_3$Si$_2$Te$_6$ and experimental setup of transient THz spectroscopy.} (a) The crystal structure of Mn$_3$Si$_2$Te$_6$. (b) Low frequency reflectivity spectrum of Mn$_3$Si$_2$Te$_6$ at room temperature. Inset: display of $R(\omega)$ up to 24000 cm$^{-1}$. The multiple sharp peaks represent infrared-active phonon modes. (c) Transformed optical conductivity of Mn$_3$Si$_2$Te$_6$ below 800 cm$^{-1}$. Inset: display of $\sigma_1(\omega)$ up to 24000 cm$^{-1}$. (d) Schematic illustration of the NIR pump-THz probe configuration. Both pump and probe beams are linearly polarized. (e) Time-domain unexcited (E$_0$) and pump-induced differential ($\Delta$E) THz electronic field. (f) Fast Fourier transform result of E$_0$ and $\Delta$E at several typical pump fluences.}\label{Fig:1}
\end{figure*}

Single crystals of Mn$_3$Si$_2$Te$_6$ were grown using solid state reaction method. High purity Mn (piece), Si (grain) and Te (lump) were taken in a stoichiometric molar ratio of 3 : 2 : 6, and placed in a quartz tube with alumina crucible under partial argon atmosphere. The sealed quartz tube for materials was heated up to 1373 K for 12 h and soaked there for another 20 h. After that, the temperature was cooled down to 1123 K at a rate of 1.5 K/h. Finally, the ampoules were taken out of furnace and the single crystals of  Mn$_3$Si$_2$Te$_6$ can be obtained. The crystal structure is shown in Fig. 1(a). The high quality (001) sample surface and ferrimagnetic transition have been verified (see Fig. S1). The static state reflectivity spectroscopy was measured by a Fourier transform infrared spectrometer (Bruker 80v). In the near-infrared (NIR) pump-THz probe experiment, the 1300 nm pump beam was generated by an optical parametric amplifier (OPA), and the THz probe beam was produced and electro-optically sampled (EOS) by 1 mm-thick ZnTe crystals (see Supporting Information). They were all based on an amplified Ti:sapphire laser system with a pulse duration of 35 fs operating at 1 kHz. The spot sizes of the NIR pump and THz probe beam at the sample surface were about 1.6 and 0.7 mm, respectively.  The terahertz optical path was filled with nitrogen to avoid absorption from water vapor (for detailed configuration, see Fig. S2). All the measurements were performed at room temperature.

\begin{figure*}[t]
	\centering
	\includegraphics[width=17cm]{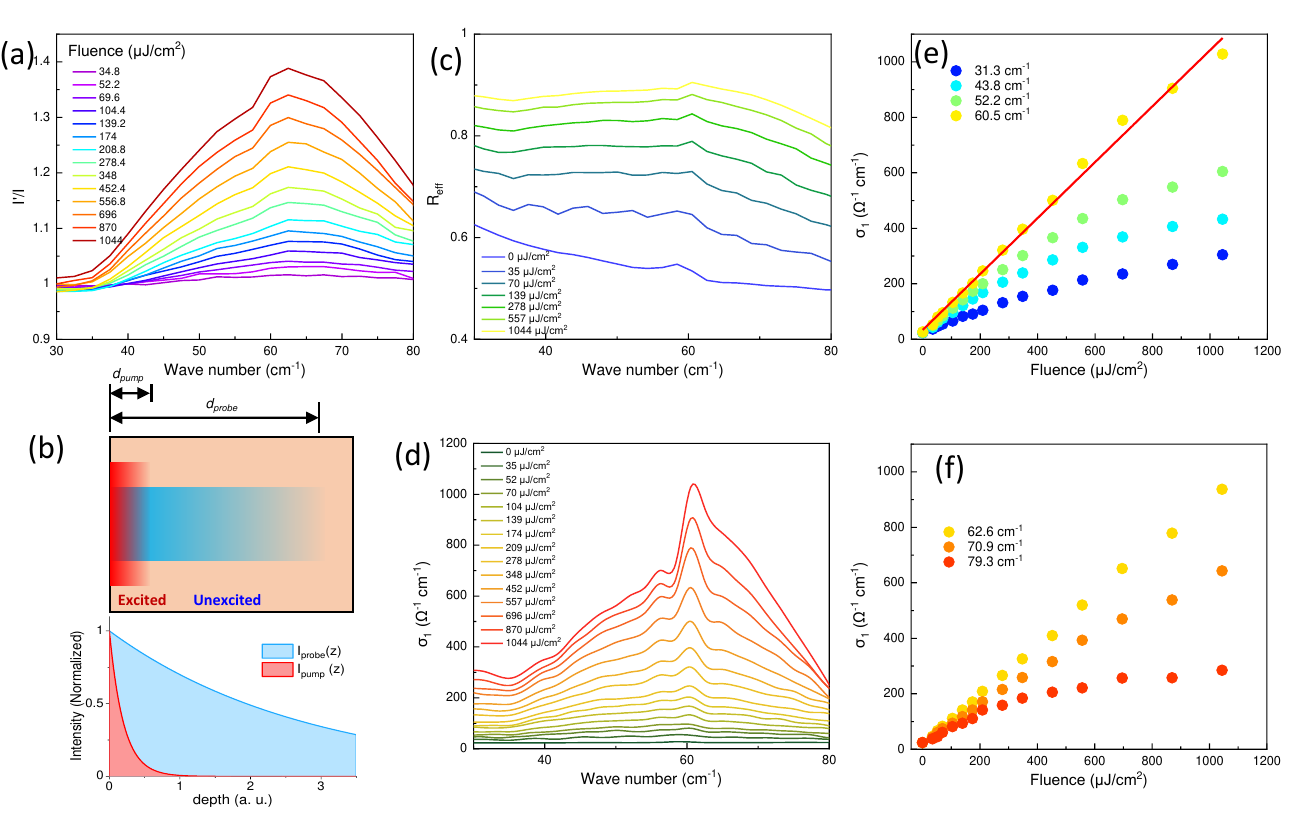}\\
	\caption{\textbf{Pump-induced optical conductivity under different pump fluences.} (a) Pump fluence dependence of diffrential THz optical intensity, which are obtained by Eq. (1). (b) Mismatch of pump and probe beams, where the skin depth of probe beam is much larger than that of pump beam. Therefore, only a vey small part of probe-covered regime (from sample surface to depth $d_{pump}$) contributes the signal. The optical response within this thin layer needs to be extracted rather than simply averaged over the entire probe-covered region. (c) The corresponding transient effective reflectivity ($R_{eff}$) spectrum (d) Extracted transient optical conductivity within surface to $d_{pump}$ by thin film mode. (e)$\&$(f) Pump fluence evolution of transient $\sigma_1$ at different frequencies. The solid line is the linear fitting result.
	}\label{Fig:2}
\end{figure*}

Figure 1(b) shows the room-temperature spectra of in-plane reflectivity $R(\omega)$ measured in Mn$_3$Si$_2$Te$_6$.  Below 400 cm$^{-1}$, $R$($\omega$) exhibits multiple sharp peaks (Fig. 1(b)), corresponding to infrared-active phonon modes. There is no signature of plasma edge at low frequency, with reflectance significantly lower than 1, indicating the extremely low carrier density. The static reflectance spectrum ($R$($\omega$)) also displays a broad hump at 15000 cm$^{-1}$, related to interband transitions. The transformed real part of optical conductivity spectrum, $\sigma_1$($\omega$), exhibits a pronounced edge below 15000 cm$^{-1}$ (inset of Fig. 1(c)), demonstrating a large-gapped band structure. Inside the absorption edge, instead of rapidly falling to a value near zero when the energy falls below the gap size, as is common with semiconductors or insulators\cite{Kim2018}, $\sigma_1$ gently decreases and the slope approaches constant at low wave numbers, this is the hallmark of indirect interband transitions. Below 400 cm$^{-1}$, the main features are multiple sharp infrared-active phonon peaks. The infrared spectra reveals that Mn$_3$Si$_2$Te$_6$ is a semiconductor with small indirect band gap, which agrees well with theoretical predictions\cite{Zhang2023}.

With NIR femtosecond laser pumping (Fig. 1(d)), the qusiparticles are instantaneously excited and lead to the transient optical response in Mn$_3$Si$_2$Te$_6$. Figure 1(e) depicts typical initial and pump-induced differential reflected electronic field, E$_0$ and $\Delta$E. Unlike typical pump-probe measurements, $\Delta$E here is even comparable in magnitude to E$_0$, indicating a significant departure from a perturbative regime. Additionally, the phase of E$_0$ does not align with that of $\Delta$E, which suggests differing frequency components. As shown in Fig.1 (f),  the spectrum of $\Delta$E exhibits a broad peak centered at 60 cm$^{-1}$, while E$_0$($\omega$) is most pronounced in the 20 to 70 cm$^{-1}$ range. This result reveals that the $E_0$($\omega$) spectra is selectively excited.

\begin{figure*}[t]
 \centering
 \centering\includegraphics[width=18cm]{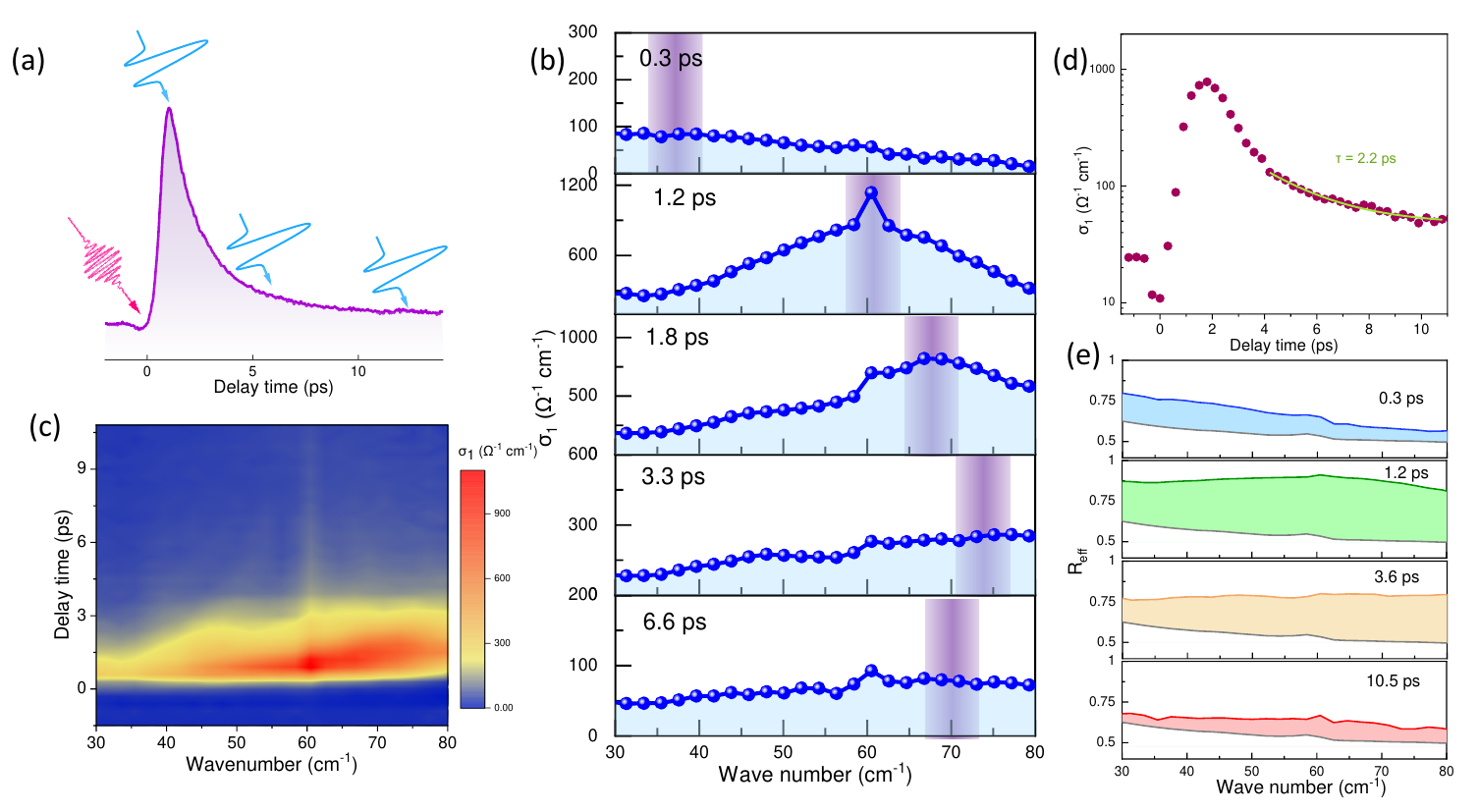}\\
 \caption{
\textbf{Delay time evolution of optical conductivity and effective reflectivity.} (a) Schematic of delay time dependent-THz reflection measurements. (b) Optical conductivity at different values of the time delay. Vertical purple bars are used to label the peak position. (c) The colorfully-mapped frequency-integrated $\sigma_1(\omega)$ response as a function of delay time. (d) The $\sigma_1$(70 cm$^{-1}$) evolution with delay time. The light green solid curve is the exponential fitting result, where the lifetime is extracted to be 2.2 ps.  (e) Illustration of $R_{eff}(\omega)$ spectrum comparing with original reflectivity at selected delay times. The gray curves represent $R(\omega)$ in static state. The colorful solid curves are transformed transient $R_{eff}$.} \label{Fig:3}
\end{figure*}

We propose the ratio of optical intensity before and after excitation, $I'/I$, to assess the enhancement of the final THz intensity:
\begin{equation}
\begin{aligned}
     \frac{I'}{I} = \frac{|E'(\omega)|^2}{|E(\omega)|^2}. \\
\end{aligned}
\end{equation}
Where $E'(\omega)$ is the fast Fourier transform result of $\Delta E(t) + E_0(t)$. Figure 2(a) depicts I'/I spectra at various pump fluences at the most significant moment of response (1.2 ps). Resembled with the line shape of $\Delta$E, a distinct broad hump emerges at 60 cm$^{-1}$ and intensifies with increasing pump fluence, reaching 1.4 at 1.04 mJ/cm$^2$. However, below 50 cm$^{-1}$, I'/I experiences a sharp decrease and eventually stabilizes near unity below 35 cm$^{-1}$. 

Note that in our experimental configuration, the skin depth of the pump beam is much smaller than that of the THz probe beam (see Fig. S3). This corresponds to a typical scenario of inhomogeneous excitation, where the exponential decay of pump beam intensity introduces depth-dependent sample properties within $d_{probe}$ (as depicted in Fig. 2(b)). To describe the optical response in the sample, we simplify the situation to a thin film model\cite{Bornoptics,Hu2014,Zhang2020}.  In this model, the sample is assumed to be uniformly excited within the skin depth of the pump beam ($d_{pump}$ = 0.25 $\mu$m, see Fig. S3), while the excitation of the deeper part of the sample can be neglected. All the transient optical constants shown below are based on the thin film model.

Then we analyze the effective reflectance of the sample after excitation ($R_{eff}$). Note that the effective reflectance is not simply the static reflectance multiplied by I'/I, it is derived with this thin film model (shown in Fig. 2(b)). As shown in Fig. 2(c), $R_{eff}(\omega)$ is overall higher than static reflectance $R(\omega)$. With increasing pump intensity, $R_{eff}(\omega)$ rises, instead of usual Drude-like response, $R_{eff}$ is enhanced more significantly near 60 cm$^{-1}$.

We observe a remarkable enhancement in optical conductivity (Fig. 2(d)). Similar to I'/I, the transient $\sigma_1$($\omega$) forms a peak at finite energy, meanwhile the near-zero energy part also has a lift, implying increasing of carrier density. It is worth noting that the gain amplitude of $\sigma_1$ is several tens of times its own, especially around 60 cm$^{-1}$, where the optical conductivity reaches 1000 $\Omega^{-1}$ cm$^{-1}$ at 1.04 mJ/cm$^2$, this is relatively rare in semiconductors. And the differential $\sigma_1$ in Mn$_3$Si$_2$Te$_6$ is even much larger than that in several typical semiconductors (see Fig. S4), such as MoS$_2$ at room temperature\cite{Strait2014}, WS$_2$\cite{BalaMuraliKrishna2018} and WSe$_2$\cite{He2018}, GaAs nanowire\cite{Parkinson2007}, CsPbBr$_3$\cite{Cinquanta2019}, and ZnSe\cite{Li2014}. The entire photoconductance spectrum is positively correlated with the excitation intensity. In Figs. 2(e)\&(f), we illustrate the relationship between the conductivity at different frequencies and the pump fluence. For any given frequency, the increase in $\sigma_1$ is both continuous and gradual. Especially, $\sigma_1$ exhibits linear growth only at 60 cm$^{-1}$ (as shown in Fig. 2(d), which can be well-fitted with a linear function), whereas the increase at other frequencies is more gradual. Both $\sigma_1$ and $R_{eff}(\omega)$ spectra indicates that the excited carriers does not simply form a Drude behavior, but a broad peak at finite energy, displaying the possbility of a new bound state in Mn$_3$Si$_2$Te$_6$.

The transient phenomena with different physical origins can usually be differentiated by analyzing the time-domain relaxation dynamics. By capturing transient THz spectra at various pump-probe delay times (Fig. 3(a)), we can determine the time-dependent optical conductivity and gain a glimpse into the underlying factors driving this unique optical response. Figure 3(b) illustrates the transient $\sigma_1$ at selected delay times. At 0.3 ps, as photons are injected, the optical conductivity begins to rise, particularly at low frequencies. The excited conductivity undergoes an immediate transformation, becoming substantially more pronounced. By 1.2 ps, the primary feature in the photoconductivity spectrum transitions into a resonant peak around 60 cm$^{-1}$. After this, the peak underwent a small blue shift and basically stabilized around 70 cm$^{-1}$ (the peak positions are labeled by vertical purple bars) combing with significant damping. The 2-dimensional colormap of $\sigma_1$ (Fig. 3(c)) provides an intuitive view of the time-dependent amplitude and frequency of $\sigma_1(\omega)$. After time zero, the blue color is converted to yellew over the entire frequency range, and rapidly changes to red in vicinity of 60 cm$^{-1}$, corresponding to the appearance of resonant state. It quickly moves to higher frequency and dissolves back towards blue color (representing equilibrium). In the whole process, the 60 cm$^{-1}$ phonon mode is survived.

We observe the relaxation rate of this instantaneous excited state by simply checking the relaxation time trace of the photoconductance at 70 cm$^{-1}$. As shown in Fig. 3(d), $\sigma_1$(70 cm$^{-1}$) experiences a rise of nearly 2 ps after time zero, increasing from 20 $\Omega^{-1}cm^{-1}$ to nearly 1000 $\Omega^{-1}cm^{-1}$, and quickly returns to the initial value. After 4 ps (while the peak position has basically stabled), the time traces exhibits a partial relaxation lifetime of approximately 2.2 ps. This relaxation is significantly faster compared to transient optical relaxation in other semiconductors\cite{Lui2014, He2018,Cinquanta2019}. From this, it can be inferred that this bound state does not immediately form after excitation but instead undergoes a 2 ps formation process and possesses a relatively short lifetime of around 2.2 ps, indicating its instability.

Similarly, the evolution of R$_{eff}$ also displays the transfer behavior of spectral weight (Fig. 3(e)). At 0.3 ps, the reflectance increase is most pronounced near zero frequency, reaching its maximum change at 60 cm$^{-1}$ by 1.2 ps. By 3.6 ps, as the reflectance begins to decrease, the point of maximum reflectance change essentially shifts to 80 cm$^{-1}$. At 10.5 ps, R$_{eff}$ has almost fully relaxed and returns to insulating state. At all delay time points, the R$_{eff}$ spectra is significantly different from the plasma edge behavior characteristic of the metallic state.

\begin{figure}[t]
 \centering
 \centering\includegraphics[width=8.5cm]{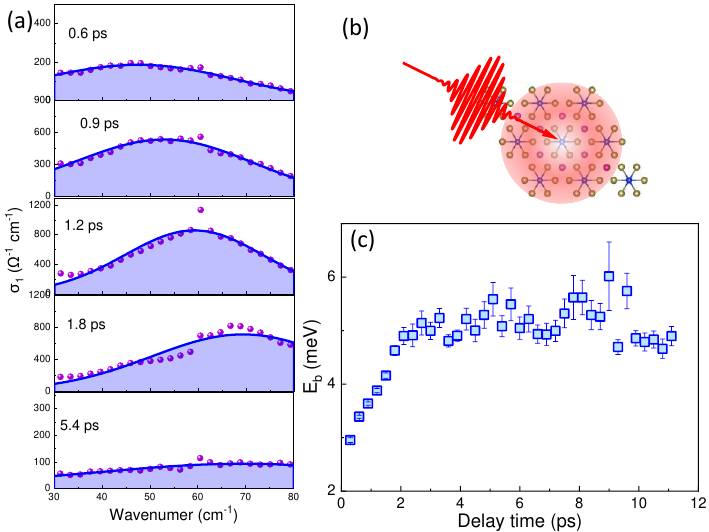}\\
 \caption{
\textbf{Description of optical conductivity spectra by polaron model.} (a) The $\sigma_1$ spectra at selected delay times fitted by a polaron model. The purple dots: experimental data points. Blue curves: fitting result by using Eq. (2). (b) The schematic of pump-induced polaron mechanism. (c) The extracted binding energy of pump-induced polaron evolving with delay time.} \label{Fig:4}                                                                                                                                                                           
\end{figure}


Many quasiparticles/excitation modes exhibit optical activity that can be distinguished based on factors such as their resonant energies and spectral line shapes\cite{Kaindl2003,Parkinson2007,Richter2010,Suzuki2012,Docherty2014,Menard2014,Xing2017,He2018,Cinquanta2019}. The THz transient spectra we measured has several noticable characteristics. Firstly, the transient $\sigma_1$ spectra shows a broad peak with non-zero dc conductivity, which deviates from common Lorentz behavior. Besides, the rapid relaxation of photoconductivity and electric field intensity (shown in Figs. 3 and S5) reflects an effective energy transfer between electrons and phonons, implying a strong electron-phonon coupling in this system. Furthermore, our experiments were carried out at room temperature, in which the low energy exciton can not be survived and further detected.  We find that the spectral change can be well explained by the pump-induced formation of polaron. Polaron describes the combination between electron and attracted lattices via electron-phonon coupling \cite{Franchini2021}, which has been experimentally observed in numerous compounds\cite{Dean2011,Suo2021}. Theoretical investigations of small-polaron response\cite{Emin1993,Emin1975} point out that the optical conductivity should have low spectral weight in the low frequency region, but a peak in the high-frequency region. The polaron response usually has the below form\cite{Wang2000}:
\begin{equation}
\begin{aligned}
     \sigma_1(\omega) = {\sigma_0}{\cdot}exp(-\frac{(\hbar\omega)^2}{\Delta^2})\cdot\frac{sinh(4E_b\hbar\omega/\Delta^2)}{4E_b\hbar\omega/\Delta^2}. \\
\end{aligned}
\end{equation}
Where $\sigma_0$ is the dc conductivity, $E_b$ is the polaron binding energy, $\Delta$ = (8$E_b$$E_{vib}$)$^{0.5}$ is a broadening factor, where $E_{vib}$ is the characteristic vibration energy. It can be seen that the polaron picture provides a high quality fit to the experimental $\sigma_1$ spectra. Taking into consideration the effective electron-phonon coupling and room temperature, it is reasonable to assign the optical response to pump-induced transient polaron (Fig. 4(b)). The fitting result by polaron picture also gives the time-dependent binding energy. As shown in Fig. 4(c), $E_b$ begins to appear from 3 meV and basically maintained around 5 meV after taking less than 3 ps. Therefore, we can provide the following simple physical picture: after optical excitation, photogenerated carriers are trapped by the lattice to form polarons within a relatively short time, accompanied by an increase in binding energy. After the binding energy stabilizes at around 5 meV, the bound carriers gradually relax back to the ground state, leading to a weakening of the polaron response. It is interesting to note that the polarons are transiently stable at room temperature even though its binding energy is small, which is different from the common equilibrium polaron. The physical origin is worthy of further study.

In summary, we observe a significant transient increase in photoconductivity and the possibly presence of pump-induced polaron in Mn$_3$Si$_2$Te$_6$ at room temperature. Following optical pulse excitation, there is a remarkable enhancement in terahertz-range photoconductivity. During the evolution of photoconductivity with delay time, the emergence of a resonant peak, accompanied by a blue-shift, is clearly visible. The transient state, once stabilized, can be well-described by a polaron model, which gives final binding energy at about 5 meV. Our work not only discovers the novel light-induced transient state in Mn$_3$Si$_2$Te$_6$ but also reveals its excellent photoelectric properties, indicating its great potential in the field of photoelectric detection.

\begin{acknowledgments}
We thank the helpful disscussion with Prof. Zhiyuan Sun and Dr. Marios Micheal. This work was supported by the National Natural Science Foundation of China (Grants Nos. 11888101 and 12174443), the National Key Research and Development Program of China (Grants Nos. 2022YFA1403901, 2018YFE0202600, and 2022YFA1403800), and Beijing Natural Science Foundation (Grant No. Z200005).
\end{acknowledgments}

\bibliography{Mn3Si2Te6-OPTP}

\end{document}